 \documentclass{aa}

\usepackage{epsfig}     
\usepackage{graphicx,color}     
\usepackage{amssymb}            
\usepackage{url}                
\usepackage{amsmath}            
\usepackage{rotating}                   
\usepackage{float}                      
\usepackage{textcomp}           
\usepackage{epstopdf}
\usepackage{dcolumn}
\usepackage{times}
\usepackage{tabularx}
\usepackage{hyperref}
\hypersetup{
    colorlinks,
    citecolor=blue,
    filecolor=blue,
    linkcolor=blue,
    urlcolor=blue,
    menucolor=black}
\usepackage{soul} 
\usepackage[english]{babel}
\usepackage{booktabs}
\usepackage{gensymb}
\usepackage{subfig}

\begin{document}


\title{ 
A highly dynamic small-scale jet in a polar coronal hole
}

\author{
Sudip Mandal\inst{1}, Lakshmi Pradeep Chitta\inst{1}, Hardi Peter\inst{1}, Sami K.~Solanki\inst{1,2}, Regina Aznar Cuadrado\inst{1}, Luca Teriaca\inst{1}, Udo~Sch\"{u}hle\inst{1}, David~Berghmans\inst{3}
\and
 Fr\`{e}d\`{e}ric~Auch\`{e}re\inst{4}}

   \institute{Max Planck Institute for Solar System Research, Justus-von-Liebig-Weg 3, 37077, G{\"o}ttingen, Germany \\
              \email{smandal.solar@gmail.com}
   \and
             School of Space Research, Kyung Hee University, Yongin, Gyeonggi 446-701, Republic of Korea
   \and
             Royal Observatory of Belgium, Ringlaan -3- Av. Circulaire, 1180 Brussels, Belgium
   \and
             Institut d'Astrophysique Spatiale, CNRS, Univ. Paris-Sud, Universit\'{e} Paris-Saclay, B\^{a}t. 121, 91405 Orsay, France
}

\abstract{
We present an observational study of the plasma dynamics at the base of a solar coronal jet, using high resolution extreme ultraviolet imaging data taken by the Extreme Ultraviolet Imager on board Solar Orbiter, and by the Atmospheric Imaging Assembly on board Solar Dynamics Observatory. We observed multiple plasma ejection events over a period of $\sim$1 hour from a dome-like base that is ca.~4 Mm wide and is embedded in a polar coronal hole. Within the dome below the jet spire, multiple plasma blobs with sizes around 1--2 Mm propagate upwards to dome apex with speeds of the order of the sound speed (ca.~120~km~s$^{-1}$ ). Upon reaching the apex, some of these blobs initiate flows with similar speeds towards the other footpoint of the dome. At the same time, high speed super-sonic outflows ($\sim$230~km~s$^{-1}$) are detected along the jet spire. These outflows as well as the intensity near the dome apex appear to be repetitive. Furthermore, during its evolution, the jet undergoes many complex morphological changes including transitions between the standard and blowout type eruption. These new observational results highlight the underlying complexity of the reconnection process that powers these jets and also provide insights into the plasma response when subjected to rapid energy injection.
}

   \titlerunning{Small-scale jet in a polar coronal hole}
   \authorrunning{Sudip Mandal et al.}
   \maketitle

\section{Introduction}

Solar jets, often described as events of collimated plasma flows, are one of the frequently observed transient features. Jets and jet-like events are not only ubiquitous throughout the solar atmosphere, but are also seen across a wide variety of spatial, temporal and temperature scales \citep{2021RSPSA.47700217S}. Jets could also play a significant role in the heating of the corona and in feeding the solar wind by supplying mass and energy \citep[e.g.][]{1983ApJ...272..329B,2000ApJ...542.1100S}.

Coronal jets (hereafter called jets for simplicity) are thought to be powered by magnetic reconnection \citep{1995Natur.375...42Y}, and are associated with events of emergence or cancellation of magnetic flux at the solar surface \cite[e.g.,][]{2012A&A...539A...7L}. Depending on the magnetic topology, jets are often observed to exhibit different morphologies and plasma dynamics \citep{2016SSRv..201....1R}. One of the classical jet morphologies is that of an inverted ``Y'' shape, also referred to as anemone jets or standard jets \citep{1992PASJ...44L.173S}. It consists of an arched dome-like base and a straight narrow spire that meets the dome at its apex. Beside this, a bright point is often observed near one of the footpoints. \citet{2010ApJ...720..757M} introduced the concept of blowout jets in which the arched core erupts violently, very often carrying cool filament material along the spire \citep{2015Natur.523..437S}. In these blowout jets, the spire is also comparatively wider than in standard jets \citep{2009SoPh..259...87N,2013ApJ...769..134M}. In some cases a standard jet has been observed to transform itself into a blowout jet \citep{2011ApJ...735L..18L}. Although most of the above mentioned studies deal with X-ray observations, their conclusions are also valid on extreme ultraviolet (EUV) jets which share a close association with X-ray jets in terms of their appearances, speeds, lifetime, or generation mechanism \citep{2007PASJ...59S.763K}.
In numerical models, a standard jet is realized as an emerging bipole reconnecting with the ambient open field \citep{1996PASJ...48..353Y,2008ApJ...673L.211M}, whereas a blowout jet is modelled as the product of either kink-instability induced interchanged reconnection \citep{2009ApJ...691...61P,2010ApJ...714.1762P} or of the magnetic-breakout mechanism \citep{2018ApJ...852...98W}. Nevertheless, details of the physical processes that occur during the triggering phase are still not understood well.

Over the years, our knowledge about coronal jets has largely been driven by studying individual jet events with high resolution, high cadence and multi-wavelength solar data. Jet dynamics such as lateral expansion \citep{2011ApJ...735L..43S}, rotating motions \citep{2007A&A...469..331J}, transverse oscillations \cite[e.g.,][]{2012ApJ...744....5M} and formation of plasmoids or blobs \cite[e.g.,][]{2017ApJ...834...79Z} are some examples of that. In fact, the detection of blobs is of particular interest as it provides insight into the magnetic reconnection process that powers a jet. In theory, when a current sheet becomes unstable, blobs can form due to the tearing mode instability \citep{1963PhFl....6..459F}. So far, detection of these blobs has mostly been limited to the jet spires \cite[e.g.,][]{2013A&A...557A.115K,2017ApJ...834...79Z,2019ApJ...870..113Z}. In our study we will show that this limitation (primarily set by the limited spatial and temporal resolution of those data) does not apply. We will present evidence of multiple plasma blobs which propagate along the jet base of a long-lived, small-scale jet region embedded within a polar coronal hole. For this we rely on EUV imaging data from the recently launched Solar Orbiter \citep[][]{2020A&A...642A...1M} spacecraft, in particular from the Extreme Ultraviolet Imager \citep[EUI;][]{2020A&A...642A...8R}. This particular EUI data set has a spatial resolution of about 420~km (2 pixels) and a cadence of 5~s. This is significantly better than earlier EUV observations covering polar regions. As a result, we are able to study how these blobs are associated with the initiation of flows along the base as well as along the jet spire. This paper is organized as follows: Section 2 describes the data wherein the results are presented in Section 3. Finally, we conclude with a discussion in Section 4.

\section{Data}

\begin{figure}[!htb]
\centering
\includegraphics[width=0.50\textwidth,clip,trim=0cm 0cm 0cm 0cm]{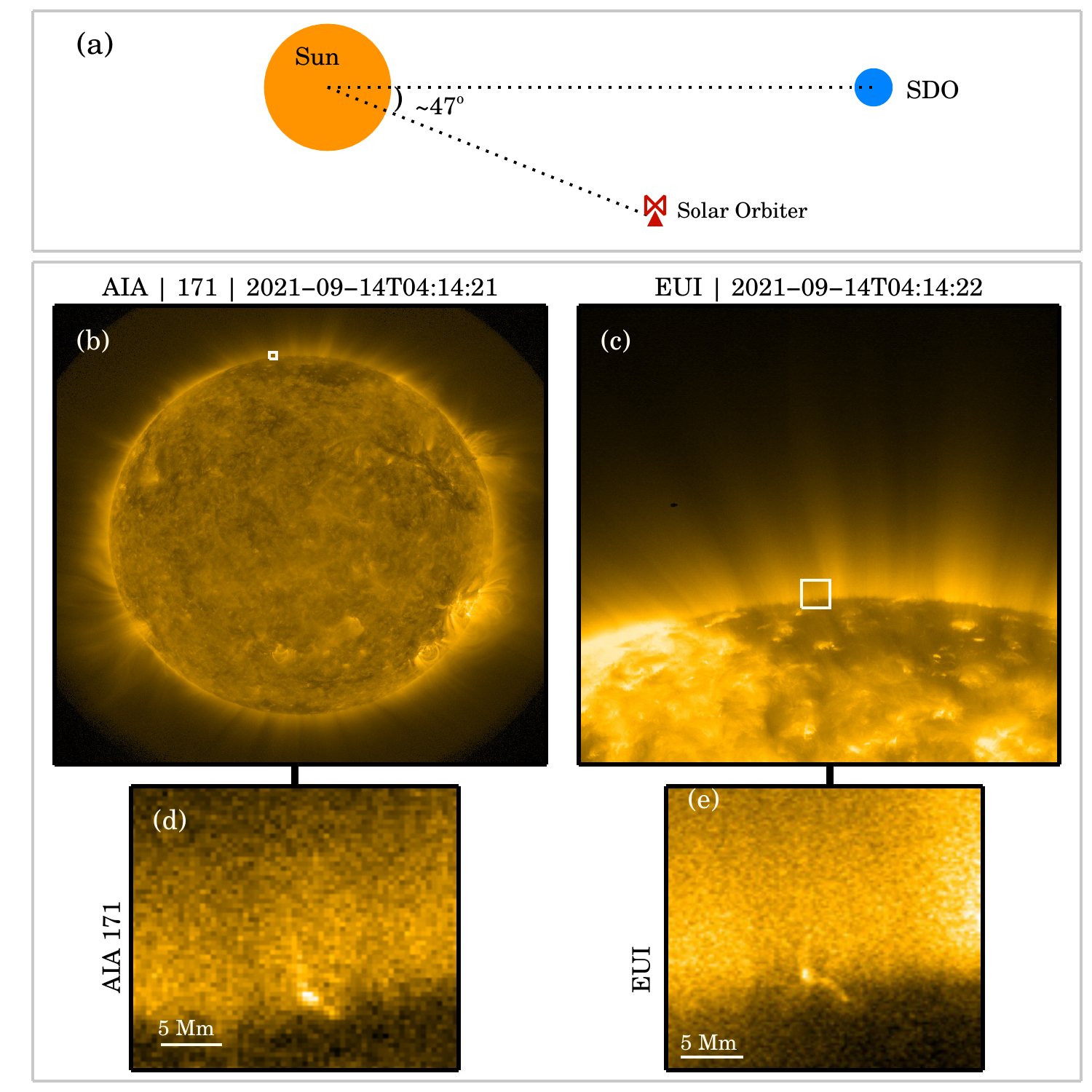}
\caption{Overview of the event. Panel-a presents a schematic diagram that outlines the relative positions of the AIA and EUI telescopes w.r.t. the Sun on 2021-Sep-14. Panel-b and Panel-c show representative context images that are taken near simultaneously by AIA and EUI, respectively. The white box in each of these images marks our region of interest (ROI) which contains the jet. Zoomed in views of the ROI are presented in Panel-d (for AIA) and in Panel-e (for EUI). An animated version of this figure is available \href{https://drive.google.com/file/d/1f1hKB29JtWGur64Fg0na2xs6sMZq-Zgl/view?usp=sharing}{here}.}
\label{fig1}
\end{figure}

In this study, we use the EUV imaging data of the north polar coronal hole taken with EUI on Solar Orbiter. In particular, we use data from the High Resolution Imager  at 174~\AA\ (HRI$_{\rm{EUV}}$) that samples plasma with a temperature of around $T\approx1$~MK. The EUI data\footnote{https://doi.org/10.24414/s5da-7e78} were obtained on 2021-Sep-14 between 04:11--04:30 UT. To extend the time covered by EUI, we also use data from the Atmospheric Imaging Assembly \cite[AIA;][]{2012SoPh..275...17L} onboard the Solar Dynamics Observatory \cite[SDO;][]{2012SoPh..275....3P}. The AIA images were taken through the 171~{\AA} (just below 1~MK) and 193~{\AA} ($T\approx1.5$~MK) passbands, between 03:20 UT and 04:40 UT on the same day\footnote{EUI observed this coronal hole region between 02:54-04:09 UT also, on the same day. However, that sequence suffers from considerable jitter and has a lower cadence of 20~s compared to the 12\,s cadence of the AIA EUV data.}. At the time of this observation, Solar Orbiter, and with it EUI, was at a distance of 0.59 AU from the Sun and its position with respect to the solar longitude was about 47{\textdegree} west from the Sun-Earth line. A schematic diagram outlining the relative position of these two spacecraft is shown in Fig~\ref{fig1}a . The plate scale of EUI in this dataset is $\sim$210 km per pixel.  For AIA this is $\sim$400 km per pixel. Moreover, the image cadence in EUI is 5~s while the AIA data cadence is 12~s. Time stamps we quote in this paper are Earth times i.e., the times measured at a distance of 1 AU, taking into account the difference in observing time due to the difference in Sun-EUI and Sun-AIA light travel times.

\section{Results}
\subsection{Event overview}

The jet that we investigate in detail occurs close to the north pole of the Sun, as highlighted by the white boxes in Fig.~\ref{fig1}b (from AIA 171~{\AA}) and \ref{fig1}c (from EUI). Common polar features such as plumes (the ray like structure stretching radially outward from the pole) as well as a coronal hole (the large darker patch) can be identified in these images. In Fig~\ref{fig1}d and e, we zoom into these boxed regions that now present a clear view of the jet structure under investigation. The overall morphology of the jet base appears to be quite similar in AIA (Fig.~\ref{fig1}d) and EUI (Fig.~\ref{fig1}e). However, EUI reveals significantly more details than AIA. This is not only due to the better spatial resolution of EUI over AIA, but could also be related to the difference in observing angle between these two instruments (Fig~\ref{fig1}a). Nevertheless, looking at the event movies (available online) from  AIA and EUI, we notice that the activity within the jet base starts around 03:29 UT and continues (rather intermittently, see the animated version of Fig~\ref{fig2}b) till 04:26 UT i.e., close to an hour in total. Therefore, the EUI observation only captured the last $\sim$20 min of this event while most of its evolution can only be seen through AIA. Hence, we first present the results using the AIA data and then discuss our findings from EUI.

\subsection{Results from AIA}
The jet event is well captured in both the 171~{\AA} and 193~{\AA} passbands and the evolutionary properties of the jets are also quite similar in these two channels. Given that the contrast of the jet base is marginally better in 171~{\AA} images, we show results from the 171~{\AA} dataset in the main text while the 193~{\AA} results are presented in the appendix.

\subsubsection{Multifaceted evolution}\label{aia_evolution}
\begin{figure*}[!htb]
\centering
\includegraphics[width=0.85\textwidth,clip,trim=0cm 0cm 0cm 0cm]{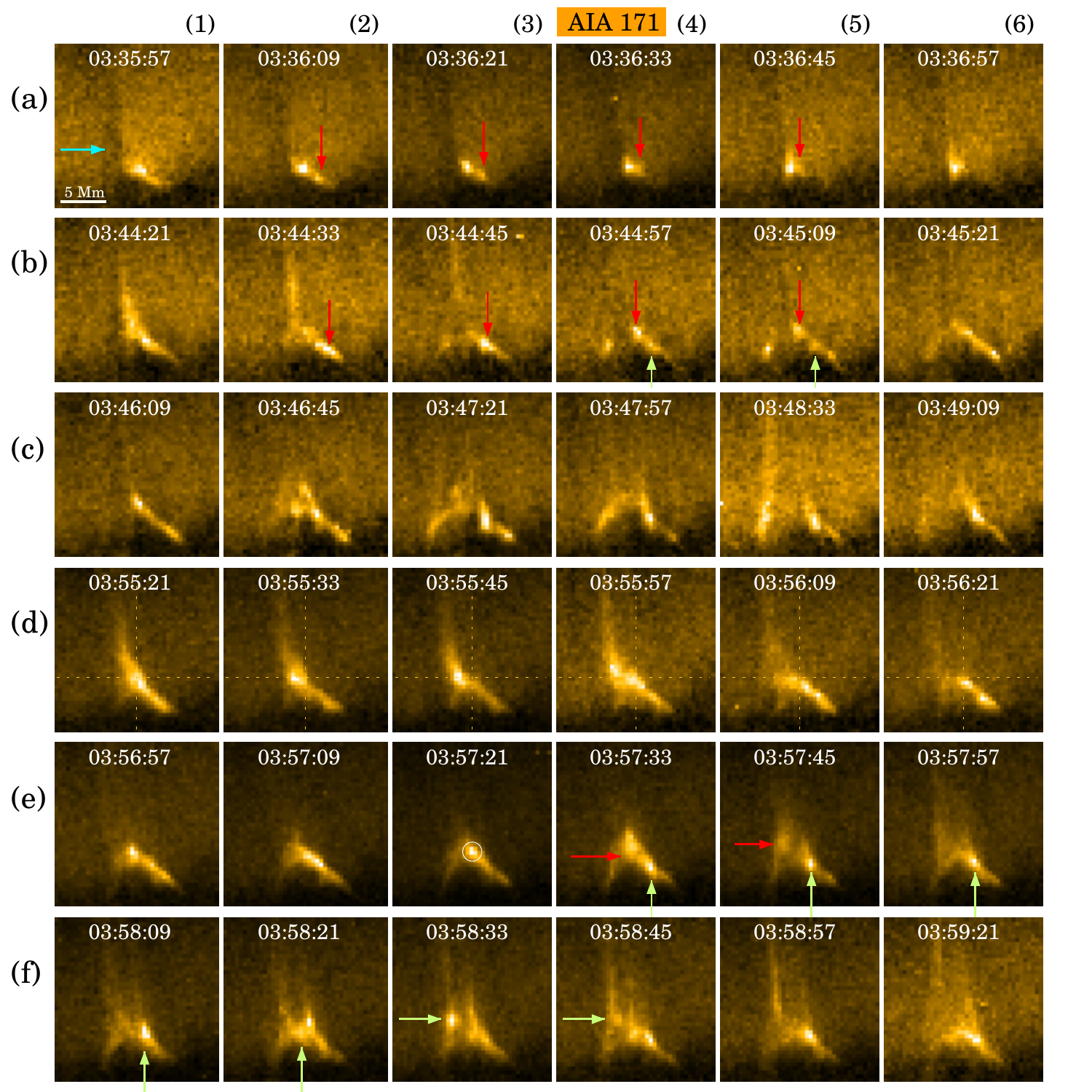}
\caption{Temporal evolution of the jet as seen in AIA 171~{\AA} images. Along each row we present a series of six AIA images (each 20$\times$18 Mm$^2$ in size) that depict a particular evolutionary stage of the jet. The circle, arrows as well as horizontal and vertical dotted lines, drawn on several panels point to a specific feature or phenomenon, as explained in Sect.~\ref{aia_evolution}. An animated version of this figure is available \href{https://drive.google.com/file/d/1giAcMwyc87-mct_t0bu3XDuxyvrGlunP/view?usp=sharing}{here}.
}
\label{fig2}
\end{figure*}

The whole event lasts for nearly an hour and over this period the jet base exhibits a multitude of dynamics such as multiple small-scale eruptions, plasma flows, and moving blobs. In Fig.~\ref{fig2} we present a summary of these evolutionary stages through a series of snapshots. We discuss each of these cases individually.

 In Fig.~\ref{fig2}a we show a total of six consecutive AIA images that contain signatures of a propagating plasma blob. Here it is very likely that the elongated spicular feature that appears darker in these images (cyan arrow) obscured a portion of the jet base. Nonetheless, we find that a blob covering about 4 AIA Pixels ($\approx$1.7\,Mm in size; highlighted by the red arrow in Fig.~\ref{fig2}a-2) first appears near the right footpoint and then travels upwards, towards another localized brightening that is stationary (for a better visualization, see the animation available online). By Fig.~\ref{fig2}a-5, we see that the blob has moved further up along the base and it is just about to merge with the stationary brightening. Certainly, the following frame (Fig.~\ref{fig2}a-6) shows an enhanced and extended brightening indicating that the merger might have happened in-between the AIA frames, in addition to any internal evolution of the stationary brightening. The blob seems to be moving slowly between Fig.~\ref{fig2}a-4 and a-5. This could be a projection effect due to the geometry of the structure. Lastly, it is possible, in principle, for successive intensity enhancements to mimic a blob-like motion (e.g., through a slipping re-connection). However, we do not think this is the case here as the stationary brightening that we observe near the apex does not exhibit any change over time (which should have been the case in response to those new reconnection events; see the animation). Hence, we interpret the observed intensity propagation in-terms of a moving blob.

Another instance of a propagating blob within this jet base is displayed in the snapshots of Fig.~\ref{fig2}b (appearing $\sim$7 min after Fig.~\ref{fig2}a-6). Firstly, the dark spicule that we previously noted in Fig.~\ref{fig2}a, has now receded, revealing a bright, thin and straight jet spire (the spire is best seen in Fig.~\ref{fig2}b-2). At the same time, we also find a brightening (marked by the red arrow in Fig.~\ref{fig2}b-2) to appear near the right footpoint of the structure. This brightening appears nearly at the same location as the previous propagating blob in Fig.~\ref{fig2}a. After 12 s (Fig.~\ref{fig2}b-3), the spire gets detached from the main structure and propagates outward, suggesting that an eruption must have occurred in-between these two AIA frames (near simultaneous AIA 94~{\AA} and 131~{\AA} images also do not show any signatures of localized heating). The propagating blob, on the other hand, continues to move upward along the arched leg. 
In the subsequent frame that is 12 s apart (Fig.~\ref{fig2}b-4), we find the spire is now completely invisible and the propagating blob has moved further up in the structure. Interestingly, another bright blob (marked by the green arrow in the same frame) appears in-between the moving blob and the right footpoint. Appearance of this new, stationary blob becomes clearer in the next frame (Fig.~\ref{fig2}b-5) whereas the previous blob (red arrow) also seems to stop moving at this point. Both these blobs loose their individual identities in the next frame (12 s later, Fig.~\ref{fig2}b-6) and we only see a uniformly bright arched base hereafter.

\begin{figure*}[!htb]
\centering
\includegraphics[width=0.79\textwidth,clip,trim=0cm 0cm 0cm 0cm]{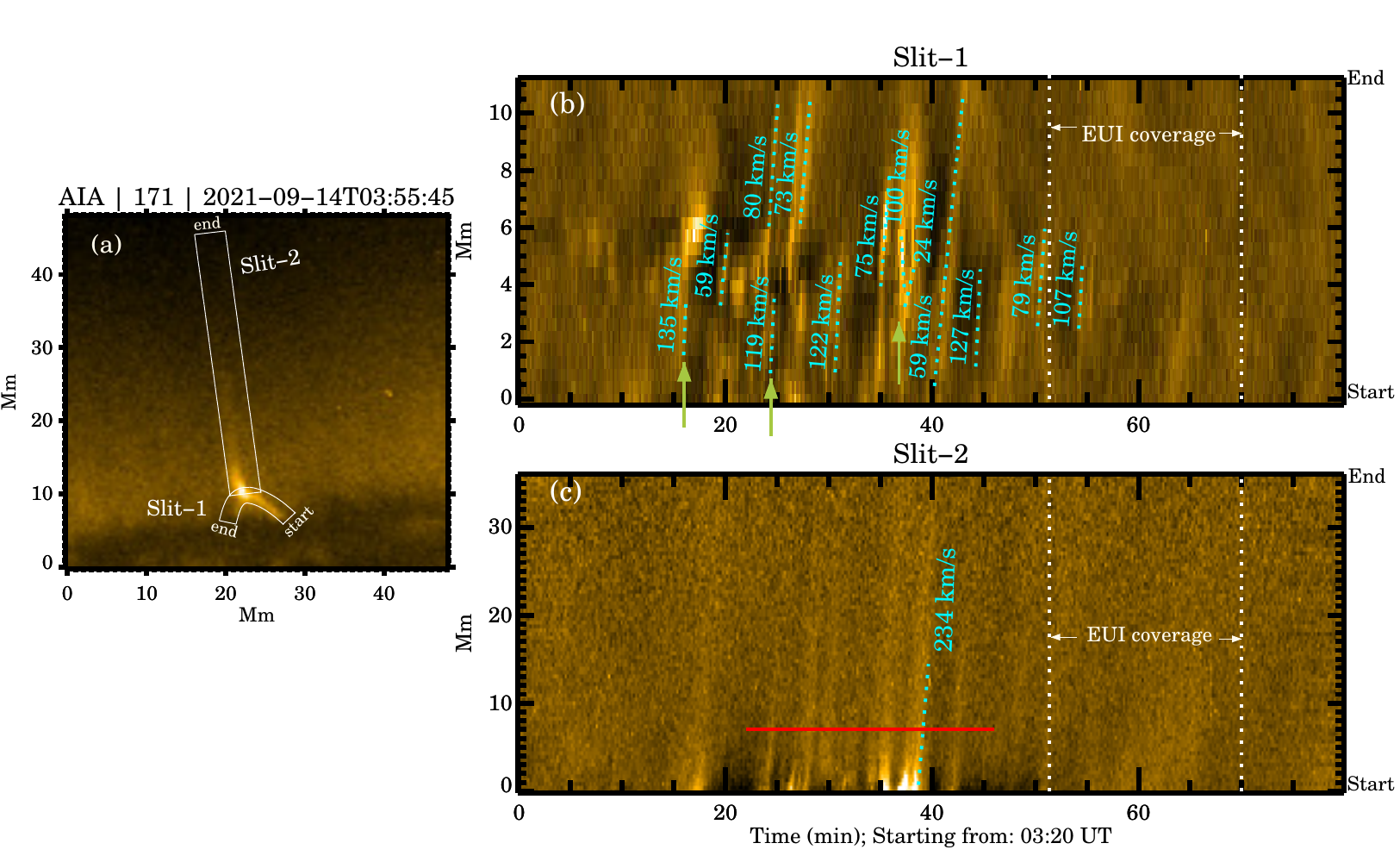}
\caption{Space-time (X-T) analysis using AIA 171~{\AA} image sequence. Panel-a shows a context image from AIA along with the two artificial slits (white boxes) that we used to derive the X-T maps. Derived maps are shown in Panel-b for Slit-1 and in Panel-c for Slit-2. In each of these maps, the slanted dotted lines (in blue) trace the inclined ridges that correspond to propagating features. Speeds calculated through the slopes of these dotted lines are printed on the panels. Furthermore, two dotted vertical lines (in white) mark the time period when EUI observations are available, wherein the red horizontal line in Panel-c marks the location at which the light curve is extracted for wavelet analysis. See Sect.~\ref{aia_xt_analysis} for further details.
}
\label{fig3}
\end{figure*}

The evolutionary stage shown in Fig.~\ref{fig2}c is rather complicated. Through these snapshots (that are now 36 s apart) we see that a significant portion of the jet base gets blown away via a series of small eruptions and at the end of it, a jet forms near the left footpoint (Fig.~\ref{fig2}c-5). This jet however does not last long and disappears within 36 sec after its inception (Fig.~\ref{fig2}c-6).

After this point the jet base again starts to brightens up slowly and the images shown in Fig.~\ref{fig2}d present a case that bears a close resemblance to the evolution of a classical jet \citep{1992PASJ...44L.173S}. Firstly, the jet has the shape of an inverted ``Y'' (or ``$\lambda$''). Over time, we see that the spire gradually brightens up and, at the same time, it moves away from the brighter right leg (Fig.~\ref{fig2}d-1,2,3). Such drifts of the spire has been noticed previously in  X-ray and EUV jets \cite[e.g.,][]{2009ApJ...702L..32S}. Regardless, the spire appears to be brightest in Fig.~\ref{fig2}d-3 and, from that point onward, it starts to fade away and almost disappears in Fig.~\ref{fig2}d-6. Interestingly, this whole evolution i.e., from spire formation to its disappearance took only a bit more than a minute.

The last two stages i.e., Fig.~\ref{fig2}e and \ref{fig2}f, depict the progression of another small eruption that happens at the jet base. At first we observe that the base expands gradually outward (Fig.~\ref{fig2}e-2,3,4) followed by an eruption (Fig.~\ref{fig2}e-5). As a result of this eruption, a jet is seen to be launched from the left base of the structure (Fig.~\ref{fig2}e-5,6). During this stage, we also notice an interesting case of counter propagating blobs which originate from a brightening near the dome apex (outlined by a circle in Fig.~\ref{fig2}e-3). One of these blobs (shown by the red arrows in Fig.~\ref{fig2}e-4,5) propagates towards the left and reach the spire before slowly fading away, wherein the other blob (highlighted by the green arrows) initially travels towards the right footpoint, before returning back towards the left (Fig.~\ref{fig2}e-4,5,6 and Fig.~\ref{fig2}f-1,2,3,4). Like the previous blob, this one too fades away slowly after reaching the spire. After this point i.e., after 03:58 UT, we do not find any coherent structure at this location other than a faint intermittent brightening near the right footpoint.

\subsubsection{AIA space-time maps }\label{aia_xt_analysis}

To capture the propagation of the observed transients, we construct space-time (X-T) maps by placing two artificial slits as shown in Fig.~\ref{fig3}. One of these slits (Slit-1) traces the jet base whereas the other slit (Slit-2) encompasses the jet spire. To increase signal, we choose the slit widths to be of 6 pixels (Slit-1) and 10 pixels (Slit-2)\footnote{As the spire moves (laterally) over the course of its evolution, we choose a rather large width of 10 pixels for Slit-2 to account for this movement.} and the derived X-T maps (averaged over their respective slit widths) are shown in Fig.~\ref{fig3}b,c. In the map of Slit-1 (Fig.~\ref{fig3}b) we find several slanted ridges of varying intensity that correspond to moving blobs and other transients that we previously identified in Fig.~\ref{fig2} (e.g., the green arrows in this X-T map point to the tracks made by some of the propagating blobs). Speeds of these transients lie in the range of 60 to 130~km~s$^{-1}$. Assuming the temperature of the emitting plasma to be about 1~MK, i.e. close to the formation temperature of the AIA 171~\AA\ band, these speeds are below or comparable to the local sound speed. 
 Before moving forward, we highlight here a couple more interesting features from this X-T map. For example, some of these transients are seen to start from the right footpoint and end near the middle (where the spire meets the dome). Others start from the middle and end at the left footpoint. 
\begin{figure*} 
\centering
\includegraphics[width=0.70\textwidth,clip,trim=0cm 0cm 0cm 0cm]{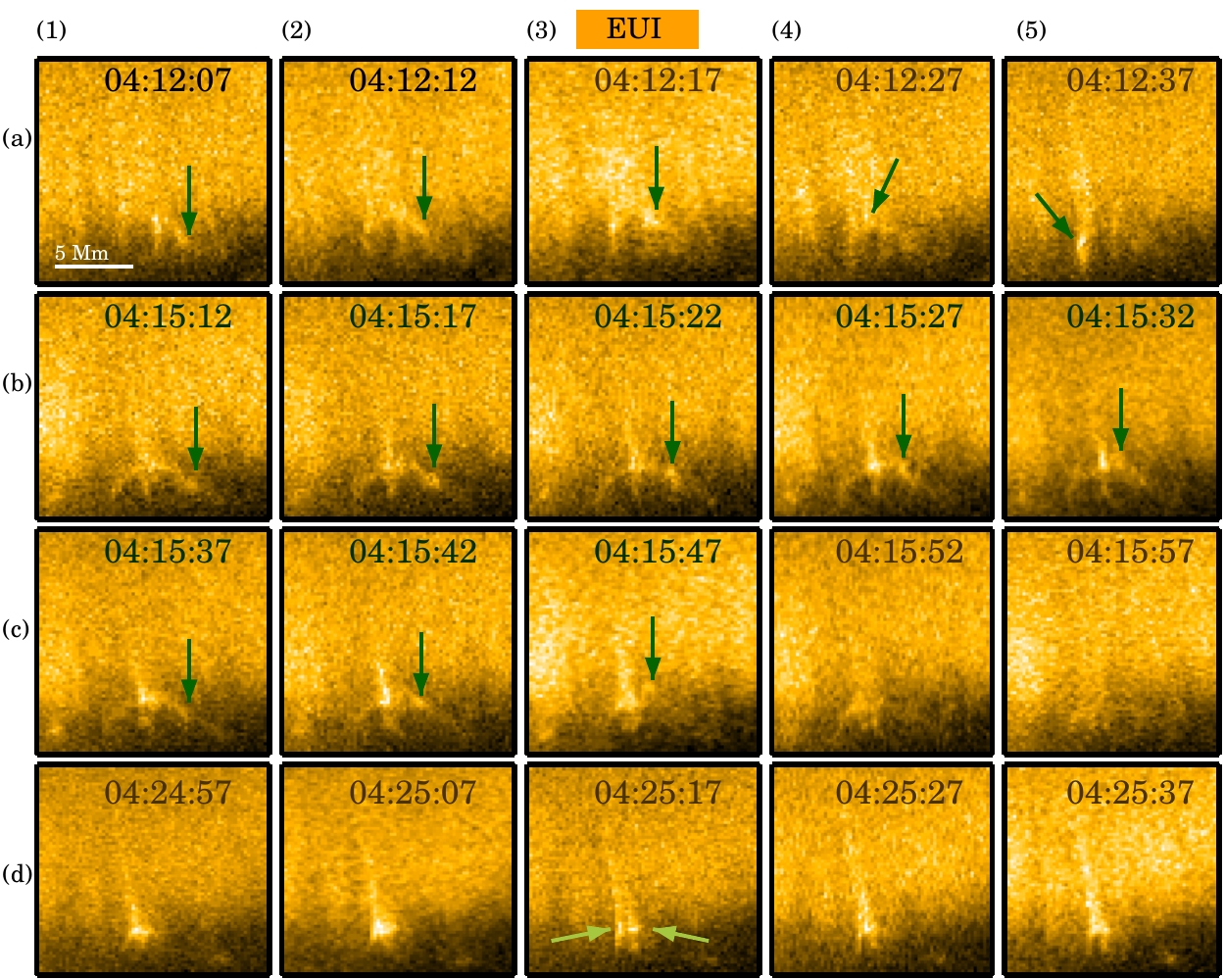}
\caption{Temporal evolution of the jet as seen through EUI images. Similar to Fig.~\ref{fig2}, each row here depicts a phase of the evolution. Arrows in some of these panels point to the instantaneous location of the propagating brightening. Each image here is of 15$\times$15 Mm$^2$ in size. An animated version of this figure is available \href{https://drive.google.com/file/d/1rpLRSwtDVCd_YiEJPt4i0k89QrkgSmkd/view?usp=sharing}{here}.}
\label{fig4}
\end{figure*}

In the Slit-2 map (Fig.~\ref{fig3}c), we find several long, regularly placed, slanted ridges that correspond to features that move along the straight spire away from the solar surface (e.g., the jets that we had highlighted in Fig.~\ref{fig2}). These ridges are only visible in the map between $t$=15 min and $t$=45 min. This is the same period within which we noted most activity in the Slit-1 map, indicating that the transients that move along the arch are casually connected with these outflows. Furthermore, these jets reach heights more than 50 Mm from their base location with speeds of $~\sim$230~km~s$^{-1}$, roughly one and a half times as fast as the local sound speed (assuming again the plasma to be of 1 MK temperature). Speeds of these features as measured from the 193~{\AA} data, are also of the same order (See Appendix~\ref{app1}). 

Lastly, a quick glance at the two AIA X-T maps reveals that during the times when EUI observations are also available (highlighted by the two vertical white lines in Fig.~\ref{fig3}b,c), the region appears to be rather quiet. Only a couple of transients are seen in the Slit-1 map (Fig.~\ref{fig3}b) whereas the Slit-2 map contains none (Fig.~\ref{fig3}c). As we will see in the next section, the EUI data do reveal a lot of dynamics, specially at the jet base, but at scales below the resolution limit of AIA.


\subsection{Results from EUI}
\subsubsection{Structural evolution  in EUI}\label{hri_evolution}
During the final 20 minutes of jet evolution, the EUI data show persistent dynamics, mostly within the jet base, which are not caught by the co-temporal AIA images. This appears to be partly because they occur at smaller spatial scales than the dynamics seen by AIA during the earlier phase of jet evolution, but also because some of the features evolve rapdily within the 12 sec cadence of AIA. Fig.~\ref{fig4} presents an overview of this.
 In Fig.~\ref{fig4}a-1, we find a bright blob-like feature (highlighted with an arrow) that shows signatures of propagation over the next couple of frames (the EUI cadence is 5~s, i.e. more than two times faster than AIA). Upon reaching the apex of the jet base, this blob drags the whole structure towards the other end of the base (Fig.~\ref{fig4}a-4; see also the event movie) and a jet forms at that location (Fig.~\ref{fig4}a-5).

The structure that we see in the snapshots displayed in  Fig.~\ref{fig4}b (and also in Fig.~\ref{fig4}c) closely resembles a fan-spine topology that consists of an inner spine, an outer spine and an arched base or dome (see Fig.1 of \citealp{2009ApJ...691...61P}). The footpoint separation between the legs of arched base we observe here is $\approxeq$ 4 Mm, which is comparable to structure extension that we noted down earlier via the AIA images (Fig.~\ref{fig2}). The next few successive frames show a bright plasma blob (of 1 Mm in size) to appear first near the right footpoint and then propagate systematically toward the dome apex. This is highlighted through the positions of the arrows in each panel of Fig.~\ref{fig4}b. Interestingly, this is followed by another propagating blob (Fig.~\ref{fig4}c-1) which originates at the same location as of the previous one and also moves towards the dome apex as seen through Fig.~\ref{fig4}c-1,2,3. Moreover, when this blob reaches the apex, the whole structure drifts leftwards (Fig.~\ref{fig4}c-4), ultimately disappearing after Fig.~\ref{fig4}c-5. Again, these blobs propagate  over a time scale of $\sim$20~s, which is the reason why they remain undetected in the 12\,s cadence images that AIA provides.

Lastly, the snapshots in Fig.~\ref{fig4}d (appearing 9 min after Fig.~\ref{fig4}c-5) reveal a similar fan-spine topology as in Fig.~\ref{fig4}c, but on a smaller spatial scale. The lateral extent of this structure is only 2 Mm, half the size of the previous one. In the first two snapshots (i.e., in Fig.~\ref{fig4}d-1,2) we only find a single brightening near the dome apex. However, Fig.~\ref{fig4}d-3 reveals that this bright feature is not a single entity but rather made up of two separate smaller brightenings (highlighted by the green arrows). In the next frame though, we again find only one brightening and it  remains unchanged in successive frames.

\begin{figure*}[!htb]
\centering
\includegraphics[width=0.75\textwidth,clip,trim=0cm 0cm 0cm 0cm]{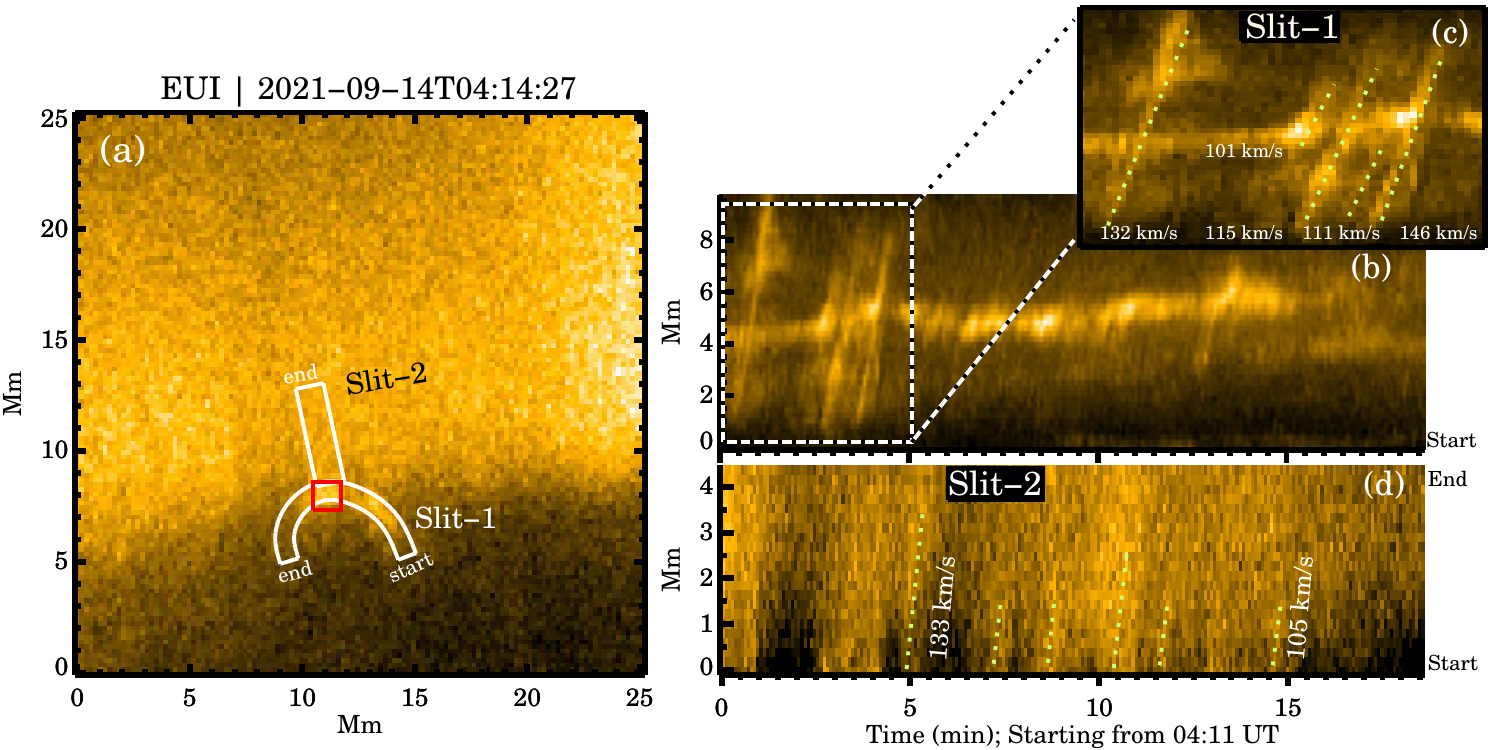}
\caption{Space-time (X-T) analysis using the EUI image sequence. Panel (a) shows a context image from EUI with the two artificial slits (white boxes) overplotted. Panel (b) presents the X-T map from Slit-1. A zoomed in view of the first 5 min of this map is shown separately as panel (c). The X-T map from Slit-2 is shown in Panel (d). Dotted lines in green trace the slanted ridges in the corresponding X-T maps and the speeds, derived from the slopes of two of these lines, are printed next to those lines. The red box in Panel-a marks the location at which the light curve is extracted for the wavelet analysis in Fig.~\ref{fig:wavelet_spectra}. See Sect.~\ref{eui_xt_analysis} for a detailed discussion.}
\label{fig5}
\end{figure*}

\subsubsection{EUI space-time maps}\label{eui_xt_analysis}

Once again, in order to measure the propagation speeds of the observed transients, we generate X-T maps using the EUI image sequence. To achieve this, we place two artificial slits (Fig.~\ref{fig5}a), one enclosing the arched base (Slit-1) whereas the other one traces the spire (Slit-2). This setup is very similar to the placement of the slits in the AIA analysis (cf. Fig.~\ref{fig3}a)
Each of these slits for the EUI analysis is 8 pixels wide and their corresponding X-T maps (averaged over the slit width) are shown in Fig.~\ref{fig5}b,c. In the Slit-1 map (Fig.~\ref{fig5}b), we find multiple slanted ridges but all within the first 5 min of the observation. This is due to the fact that after this time, the arch structure disintegrates (Fig.~\ref{fig4}c-5) and hence, no propagation  can be found. The estimated speeds of these features are between 101~km~s$^{-1}$ and 146~km~s$^{-1}$ (Fig.~\ref{fig5}c), comparable to the local sound speed if we assume the plasma to be at $\sim$1 MK temperature.

An interesting feature to note here is that three out of these five ridges extend from one end to the other end of the space axis, i.e. they stretch along the whole length of Slit-1. However, these are not due to the blobs themselves making an end-to-end propagation along Slit-1, but rather due to the blobs triggering flows after reaching the dome apex (see also the EUI movie). It is then imperative to ask whether or not these blobs initiate similar flows along the spire when they reach the dome apex. In Fig.~\ref{fig5}d we show the X-T map from Slit-2 which traces those transients that move along the spire. The signals are  weaker in this case and hence, only a few ridges can be identified with conviction. Speeds of these transients lie between 109~km~s$^{-1}$ and 136~km~s$^{-1}$. Although these values are fairly comparable to the blob speeds, they are significantly less than the outflow speeds that we had measured earlier from the AIA data. Further, the association between a propagating blob in Slit-1 (Fig.~\ref{fig5}b) and a transient seen in Slit-2 (Fig.~\ref{fig5}d) cannot be established conclusively due to weaker signal in the Slit-2 map. Nonetheless, these outflow speeds are significantly higher compared to the speeds of micro-jet like events that were reported by \citet{2021ApJ...918L..20H} using similar or coarser spatio-temporal EUI data covering a quiet-Sun region \footnote{For reference, most of the micro-jets found by \citet{2021ApJ...918L..20H} have speeds approx. 60~km~s$^{-1}$ with very few reaching upto 180~km~s$^{-1}$. Moreover, all those micro-jetting cases were on-disc events and their true speeds may well be even lower.}. Therefore, the higher flow speeds that we observed here could be facilitated by ambient open magnetic fields in the coronal hole region.

\subsection{Repetitive occurrences } \label{periods_in_data}

\begin{figure*}[!htb]
\centering
 \captionsetup[subfigure]{labelformat=empty}
  \subfloat[ ]{\includegraphics[trim = 0cm 0cm 0cm 0cm, clip,width=0.49\textwidth]{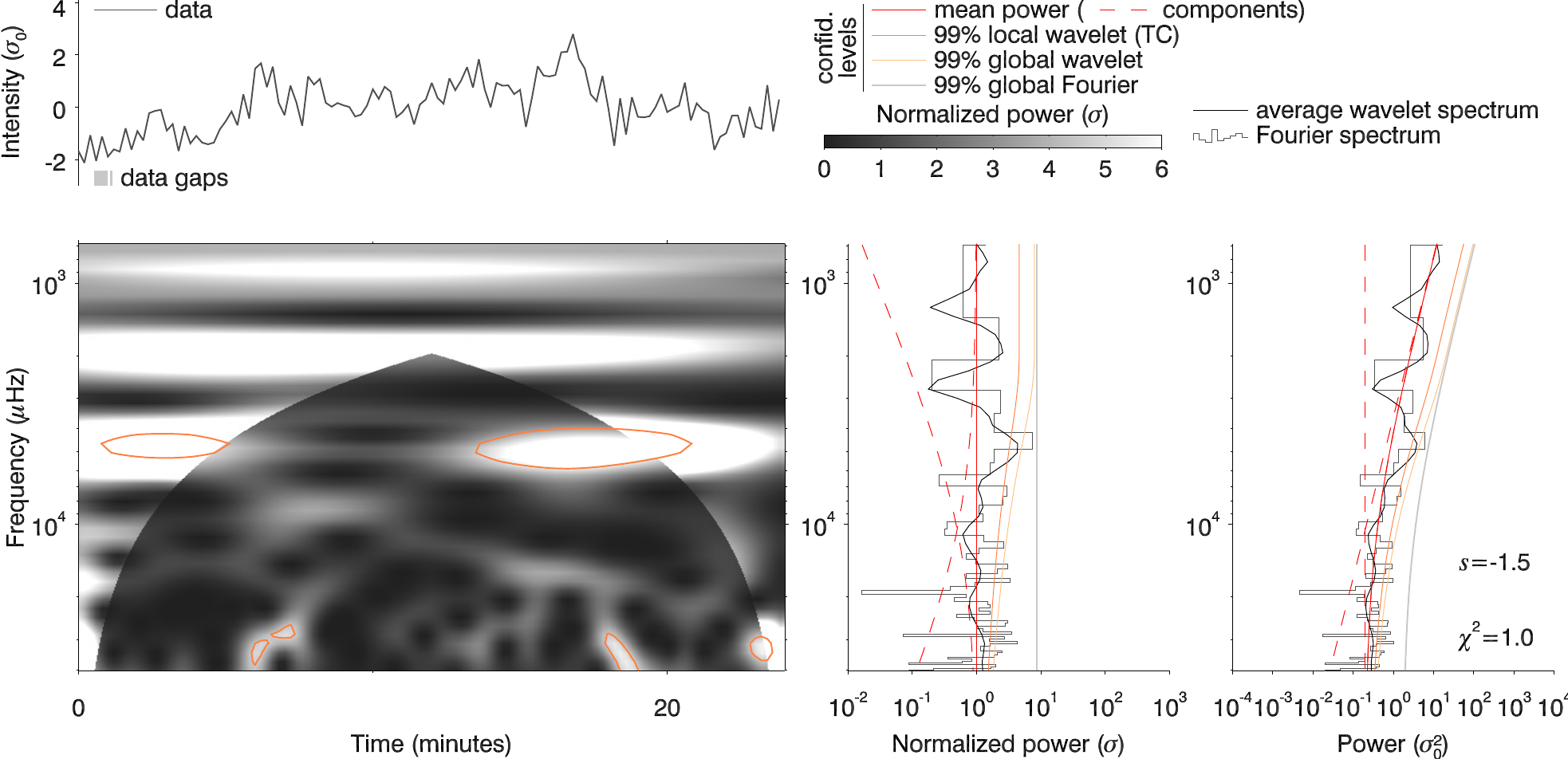}}
  \subfloat[ ]{\includegraphics[trim = 0cm 0cm 0cm 0cm, clip,width=0.49\textwidth]{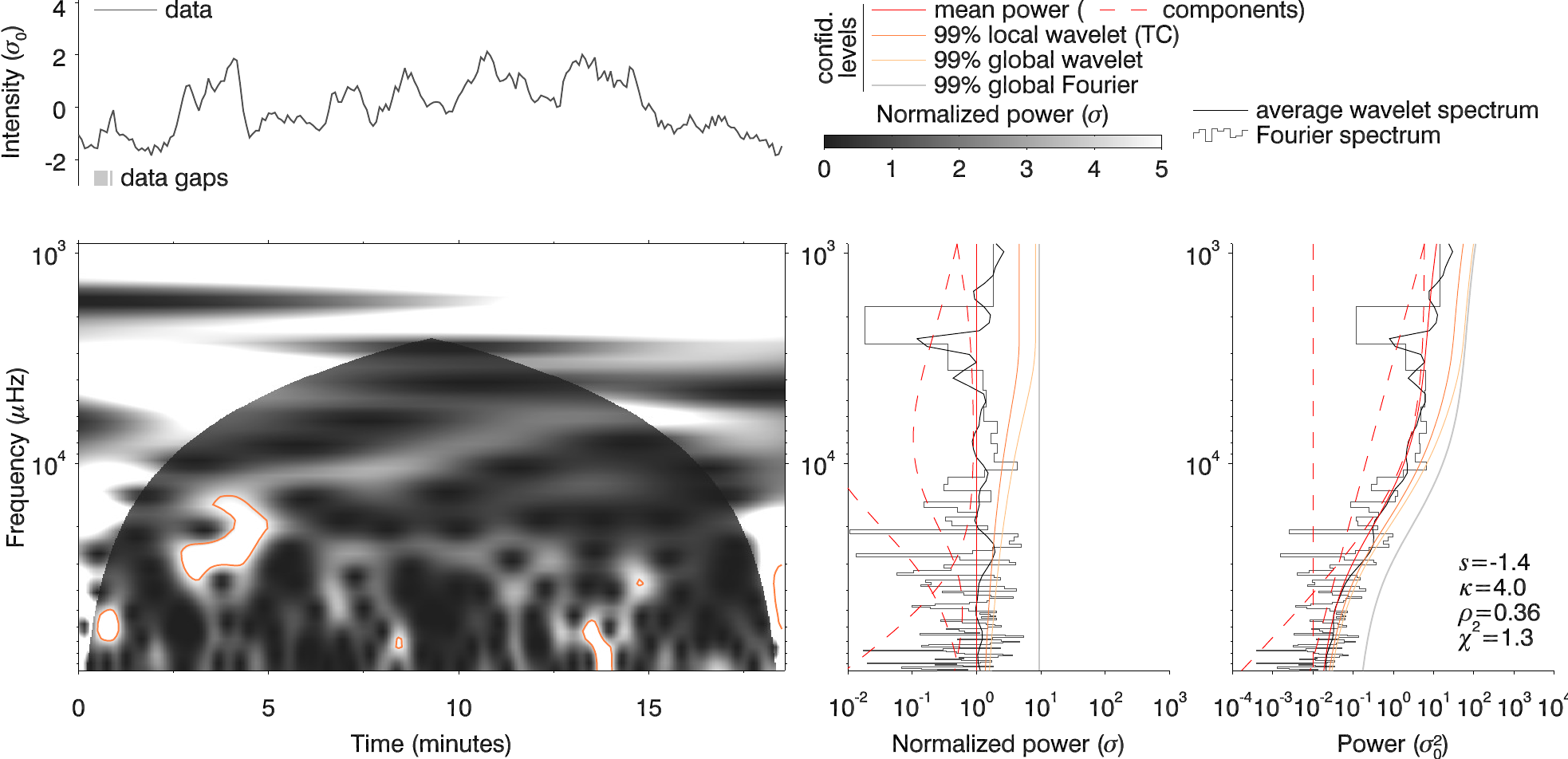}}
\caption{Wavelet and Fourier analysis of the temporal evolution for AIA (left panel) and EUI (right panel).  For each panel, the variance-normalized light curve is shown at the top left. Below is the whitened Morlet wavelet spectrum (i.e. normalized to the model of background power). The right and size plots show the time-averaged wavelet (thick line) and Fourier (histogram) power spectra. The fitted model of background power is plotted in red, along with its components (dashed red). The dark and light orange curves respectively give the 99\% local and global confidence levels for the time-averaged wavelet spectrum. The gray line represents the 99\% global confidence level for the Fourier spectrum. In each panel, the middle plot represents the same information with whithened spectra.}
\label{fig:wavelet_spectra}
\end{figure*}

In Sect.~\ref{aia_xt_analysis}, we noted that the flows that move along the jet spire, occur at regular intervals (See Fig.~\ref{fig3}b). In order to probe this further, we first extract a light curve from the Slit-1 map of AIA 171~{\AA} (the location is shown by the red horizontal line in Fig.~\ref{fig3}b). We restrict the curve to between t=21 min and t=45 min only, i.e., to the times when these outflows appear in the map. We perform a wavelet and Fourier analysis of these two light curves following the methodology introduced by ~\cite{2016ApJ...825..110A}. The wavelet analysis uses the code by \citet{1998BAMS...79...61T} with a custom noise model. The power spectrum of the AIA time series can be modelled by a simple power-law plus white noise:
\begin{equation}
\sigma(\nu)=A\nu^s+C.
\label{eq:power_model}
\end{equation}

where $\nu$ is the frequency, s is the power index while A and C are constants. The results are plotted in the left panels of Fig.\ref{fig:wavelet_spectra}. Before discussing the results, let us also use the EUI data. Now, a similar analysis with the Slit-2 map of EUI is not possible due to poor signal strengths. To overcome this, we select a 10$\times$10 pixel region near the jet base (the red box in Fig.~\ref{fig5}a) and extract a box-averaged light curve from there (top right panel of Fig.~\ref{fig:wavelet_spectra}). It is quite evident from the light curve that the intensity does wax and wane with time.
As can be seen in the right plot of the right panel of Figure~\ref{fig:wavelet_spectra}, the background power spectrum of this time series cannot be modelled by a power law as was the case for the AIA time series. It exhibits a change of slope that forms a hump around 7mHz (2.4 min), which may correspond to the above-mentioned recurring fluctuations. In this case, we use the following model;
\begin{equation}
\sigma(\nu)=A\nu^s+\text{K}_\rho(\nu)=B\left(1+\frac{\nu^2}{\kappa\rho^2}\right)^{-\frac{\kappa+1}{2}}+C
\end{equation}
where $\text{K}_\rho(\nu)$ is the Kappa function in which $\rho$ is the width of the PSD (power spectral distribution) and $\kappa$ defines the extent of its high-frequency wing (see \citealp{2016ApJ...825..110A} for further details).

The additional term compared to Equation~\ref{eq:power_model} was introduced by \cite{2016ApJ...825..110A, 2016ApJ...827..152A} to model the power spectral distribution of time series containing periodic pulses of random amplitudes. However, in the present case, the PSD does not exhibit the discrete peak characteristic of a periodic signal (right panel of Figure~\ref{fig:wavelet_spectra}). Overall, although the light curves from AIA and EUI show recurring fluctuations, no significant power is found at the 99\% confidence level (the peak at 3.4 min in AIA series lies just below the confidence interval). Hence, we conclude that the intensity near the dome apex (where the jet spire meets the jet dome) changes repetitively (but not periodically) and it leads to recurring outflows along the spire.

\section{Discussion and conclusion}

In this paper, we presented the evolution of a small-scale, long-living structure that repeatedly ejected jets. This feature was located within an inter-plume region embedded in a polar coronal hole. Considering the observed evolution in EUI as well as in AIA data, it is quite clear that the jet base went through multiple reconnections over a time span of an hour. We could imagine (at least) two ways how this could happen. In Method-1, continuous flux emergence keeps the emerging field reconnecting with the pre-existing ambient field \citep{1996PASJ...48..353Y}. Since the jetting region in this event is located close to the limb, the underlying photospheric magnetic field information, e.g., from a magnetogram is unfortunately not available. Still, evidence of such episodic flux emergence in connection to jets, albeit for an active region, has previously been reported by \citet{2019ApJ...883...52R}. The other mechanism that can trigger a jet, Method-2, involves a mini-filament eruption \citep{2015Natur.523..437S}. A close look at the AIA event movie (between 3:28:16 UT and 3:29:28 UT) reveals dark material moving away from the jet base during the first jet. This could well be the cool mini-filament material or at least a part of it. However, the jet base is mostly covered by the spicular forest and hence, we could not directly identify a (low-lying) filament before the eruption. Nevertheless, we suggest that it is most likely that over the course of this event, both these triggering mechanisms could have played their part. For example, the top of the jet base getting completely blown away in Fig.~\ref{fig2}c-5 is probably an indication of Method-2, whereas the fact that at a later time in Fig.~\ref{fig2}d-4, we find a classical jet like structure gradually appearing (and disappearing) at the same location, is probably indicative of Method-1. In this latter case, magnetic flux might have newly emerged at the surface after the eruption of the mini-filament.

Another important aspect of this event is the presence of propagating blobs. Using EUI data, \citet{2021A&A...656L..16M} reported several blob-like brightenings (campfire blobs) in the quiet Sun corona that show clear signatures of propagation but without any associated jets or outflows. Although in one of those cases, campfire blobs were seen to be repeatedly appearing from a given location, their propagation speeds were significantly lower ($\sim$30~km~s$^{-1}$; well below the sound speed) than in the jet events we discuss in this study, where the speeds reach the sound speed (using data from Hi-C, \citealp{2015ApJ...801L...2P} also reported flows of similar speeds, albeit from a different magnetic configuration). Moreover, it is believed that such campfire blobs are generated through magnetic reconnection between low lying loop like structures \citep{2021A&A...656L...4B}.
3D MHD models of the jet formation \cite[e.g. by][]{2017ApJ...841...27N} have already demonstrated that in low-plasma $\beta$ situations, blobs can easily form due to reconnection dynamics and propagate upwards in jets. However in our case, the propagating blobs are mostly restricted to remain along the jet base. Furthermore, the blobs originate more or less from the same location of the base (slightly above the right footpoint, Fig~\ref{fig4}b-1,c-1) and no signatures of simultaneous counter-propagating blobs are detected (except the one case seen in Fig.~\ref{fig2}e). These considerations suggest that (a) the initial reconnection site is hidden under the spicular forest and, (b) these blobs are not energetic enough to retain their individual identities beyond the dome apex. In fact, some of our observed blob properties share a close resemblance with the simulation results of \citet{2013ApJ...777...16Y}, who modelled blob-like feature seen in chromospheric anemone jets \citep{2012ApJ...759...33S}. Just like in our case, these authors found that the modelled blobs often get infused into the jet as they reach the dome apex. However, the aforementioned simulation was set in a relatively high-plasma $\beta$ situation and hence, other effects such as the Kelvin-Helmholtz instability may have had an effect on the evolution of those blobs \citep{2017ApJ...841...27N}.

Finally, we consider the recurring nature of the intensity fluctuations that we observed near the dome apex and subsequently in the outflows. Given the peak around 3.4 min that we found in the AIA time series  (although it was not significant at 99\% confidence level), it is tempting to consider the scenarios of wave modulated reconnection \citep{2006SoPh..238..313C} or an oscillatory type reconnection \citep{1991ApJ...371L..41C}. In fact, \citet{2009A&A...494..329M} showed how emerging flux on the backdrop of a pre-existing open ambient field can give rise to oscillatory reconnection events that are capable of generating vertical upflows with periods between 1.75-3.5 min \citep[see also][]{2012ApJ...749...30M}. Unfortunately, as we found out previously, there is no significant power at the 99\% confidence level in either of our time series and hence, we can not make any firm conclusion based on these evidences.

 To conclude, by analysing the highest resolution EUV observation of a polar coronal jet observed at the time the data were taken, we found that recurring jets can exhibit vastly different evolution from one another, including an atypical morphological transition from a standard to a blowout jet and back. Some of these changes are partly modulated by plasmoids which are by-products of magnetic reconnection that powers these jets in the first place. Other processes such as flux emergence also play a part driving such an evolution.

\section{Acknowledgments}
We thank the anonymous reviewer for the encouraging comments and helpful suggestions. Solar Orbiter is a mission of international cooperation between ESA and NASA, operated by ESA. The EUI instrument was built by CSL, IAS, MPS, MSSL/UCL, PMOD/WRC, ROB, LCF/IO with funding from the Belgian Federal Science Policy Office (BELSPO/PRODEX PEA 4000134088); the Centre National d’Etudes Spatiales (CNES); the UK Space Agency (UKSA); the Bundesministerium f\"{u}r Wirtschaft und Energie (BMWi) through the Deutsches Zentrum f\"{u}r Luft- und Raumfahrt (DLR); and the Swiss Space Office (SSO). The AIA data used here is courtesy of the SDO
(NASA) and AIA consortium. The authors would also like to acknowledge the Joint Science Operations Center (JSOC) for providing AIA data download links.

 \bibliographystyle{aa}
\bibliography{ref_sep14_north_pole_jet}

\newpage

\section{Appendix}

\subsection{AIA 193~{\AA} space-time (X-T) map}\label{app1}
 
\begin{figure}[!htbp]
\centering
\includegraphics[width=0.49\textwidth,clip,trim=0cm 0cm 0cm 0cm]{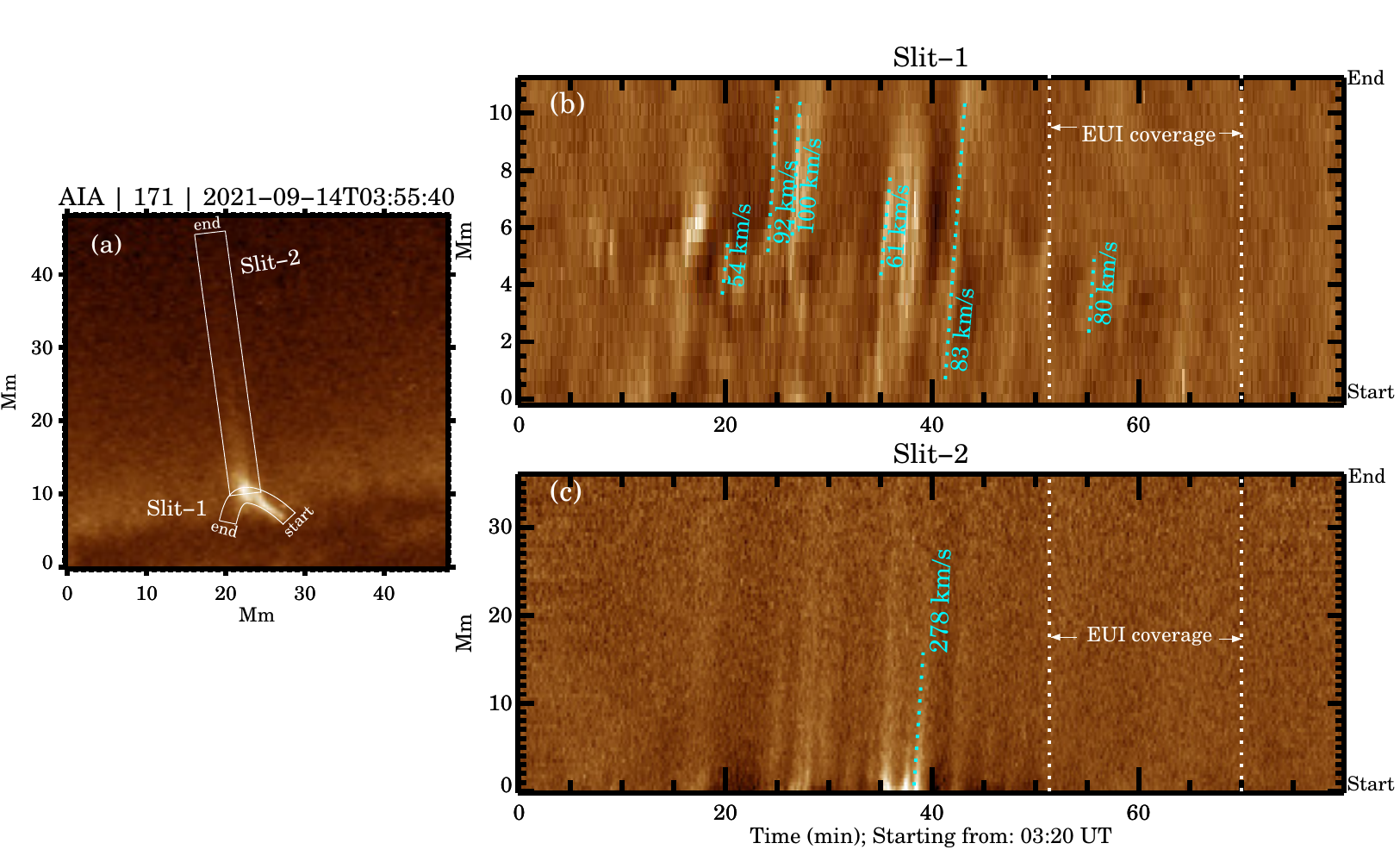}
\caption{Space-time analysis using AIA 193~{\AA} image sequence. Panel descriptions are same as Fig.~\ref{fig3}.}
\label{app_xt_193}
\end{figure}

As mentioned in the main text, the evolution of the jet in the AIA 193~{\AA} channel is quite similar to that of the 171~{\AA} channel. In Fig.~\ref{app_xt_193} we present two X-T maps (Panel-b,c) generated using 193~{\AA} data. These maps not only show features similar to Fig~\ref{fig3}, but also reveal that their propagation speeds are comparable.

\subsection{Co-temporal AIA-EUI snapshots}

\begin{figure}[!htb]
\centering
\includegraphics[width=0.49\textwidth,clip,trim=0cm 0cm 0cm 0cm]{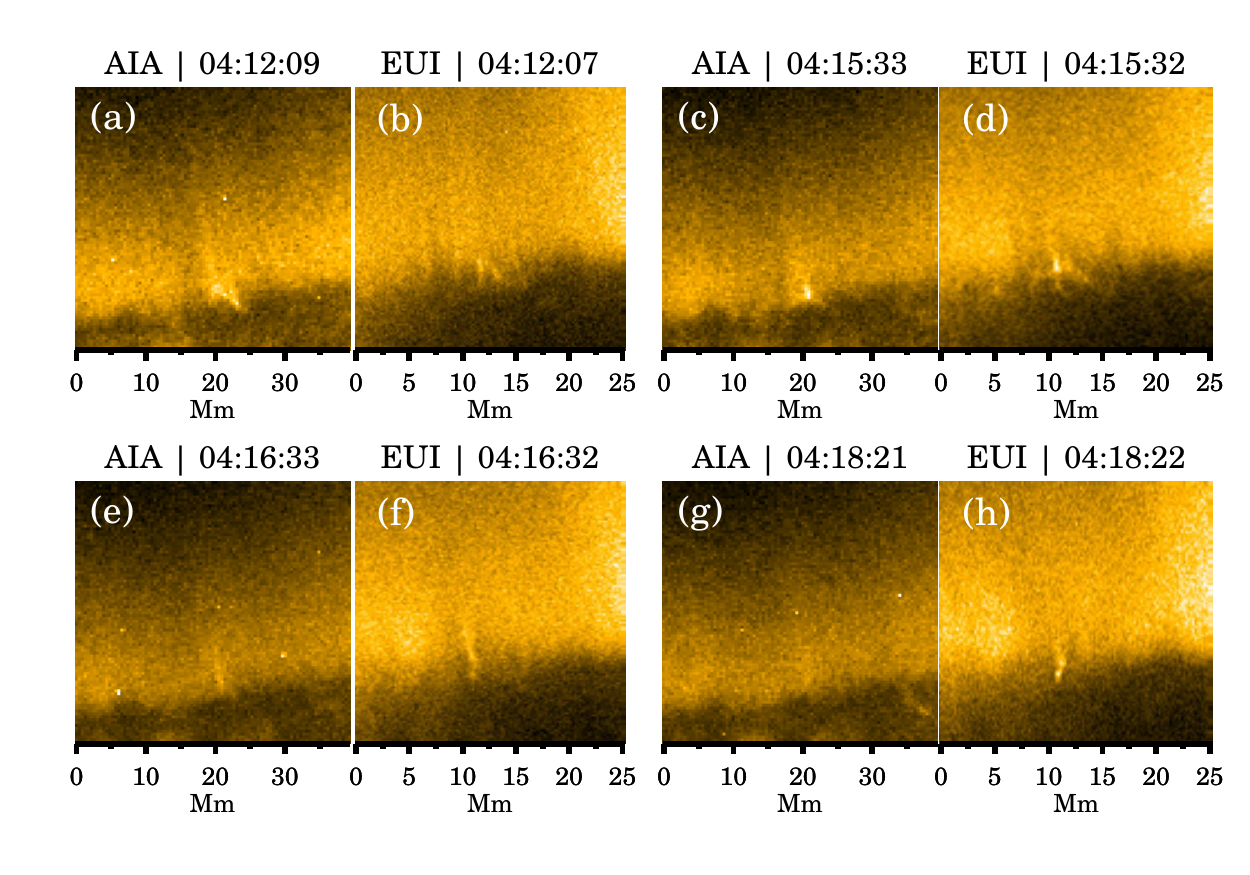}
\caption{Examples of co-temporal images obtained with AIA 171~{\AA} channel and EUI.}
\label{app_aia_eui}
\end{figure}

Here we present a few more examples of co-temporal images from the AIA 171~{\AA} channel and EUI (Fig~\ref{app_aia_eui}). These images underline the better spatial resolution of EUI when compared to AIA.   

\end{document}